\documentclass[12pt]{article} 
\usepackage[T2A]{fontenc}
\usepackage{graphicx}
\usepackage[cp1251]{inputenc}
\begin{document}

\def\physscr{Physica Scripta}
\def\aap{Astron. Astrophys.}
\def\mnras{MNRAS}
\def\aj{Astron. J.}
\def\aaps{Astron. Astrophys. Suppl. Ser.}
\def\apss{Astrophys. Space Sci.}
\def\arxivprefixe{arXiv}
\def\arxivprefixesep{:}
\newcommand{\eprint}[2][]{%
	{\tt\if!#1!#2\else#1\arxivprefixesep\ignorespaces#2\fi}%
}


\newcommand{\Teff}{T_{\rm eff}}
\newcommand{\lgg}{\log\,g}
\newcommand{\eps}[1]{\log\varepsilon_{\rm #1}}
\newcommand{\kms}{{\rm km/c}}
\newcommand{\kH}{S_{\!\rm H}}    
\newcommand{\Eexc}{$E_{\rm exc}$}
\newcommand{\iso}[1]{\mbox{$^{#1}{\rm Ba}$}}
\newcommand{\eu}[5]{\mbox{$#1\,^#2{\rm #3}^{#4}_{\rm #5}$}}

\centerline{\bf Abundances of $\alpha$-Process Elements}
\centerline{\bf in Thin-Disk, Thick-Disk, and Halo Stars of the Galaxy: Non-LTE Analysis}

\bigskip
\centerline{Lyudmila Mashonkina$^1$\footnote{\tt E-mail: lima@inasan.ru}, Maria Neretina$^{1,2}$, Tatyana Sitnova$^1$, Yuri Pakhomov$^1$}

\bigskip
\centerline{\it $^1$Institute of Astronomy, Russian Academy of Sciences,}
\centerline{\it Pyatnitskaya st. 48, 119017 Moscow, Russia}
 
\centerline{\it $^{2}$Moscow State University, Vorob'evy gory, Moscow, 119899 Russia}
 
\bigskip
{\bf Abstract -}
The atmospheric parameters and abundances of Mg, Si, Ca, and Ti have been determined for 20 stars using the Gaia DR2 parallaxes, high-resolution spectra, and the non-local thermodynamic equilibrium (non-LTE) line formation modeling. A sample of stars with homogeneous data on the abundances of $\alpha$-process
elements has thus been increased to 94. It is shown that applying a non-LTE approach and classical 1D
atmospheric models with spectroscopically determined surface gravities based on Fe~I and Fe~II lines yields reliable results. Analysis of the full sample confirms the conclusions of earlier studies
indicating enhancements of Mg, Si, Ca, and Ti relative to Fe for halo and thick-disk stars, and larger enhancements for the thick-disk stars compared to the thin-disk stars of similar metallicities. The following new results are obtained. The ratios [Mg/Fe], [Si/Fe], [Ca/Fe], and [Ti/Fe] in the thick disk remain constant and similar to each other at the level 0.3 when [Fe/H] $\le -0.4$, and fall off when the metallicity becomes higher, suggesting the onset of the production of iron in Type Ia supernovae. Halo stars have the same [$\alpha$/Fe] values independent of their distance (within $\sim$ 8~kpc of the Sun), providing evidence for a universal evolution of the abundances of $\alpha$-process elements in different parts of the Galaxy. The enhancements relative to iron for halo stars are, on average, similar, at the level 0.3~dex, for Mg, Si, Ca, and Ti. These data are important for constraining the nucleosynthesis models. 
The star-to-star scatter of [$\alpha$/Fe] increases for [Fe/H] $\le -2.6$, while the scatter of the ratios between the different $\alpha$-process elements remains small, possibly indicating incomplete mixing of nucleosynthesis products at the epoch of the formation of these stars.



\section{Introduction}

There remain many unanswered questions about the formation and evolution of our Galaxy, one of which is the formation of the thick disk, which was discovered comparatively recently in an analysis of the distribution of the stellar density perpendicular to the Galactic plane \cite{1983MNRAS.202.1025G}. Studies of the chemical composition of the thick disk are important for understanding the origin of its stellar population. Gratton et al. \cite{1996ASPC...92..307G} and Fuhrmann \cite{Fuhrmann1998} found that, in the range $-1 \le$ [Fe/H]\footnote{We use standard notation for abundance ratios: 
[X/Y] = log$(N_X/N_Y)_{*}$  -- log$(N_X/N_Y)_{\odot}$.} $\le -0.3$, stars of the thick disk display enhancements of oxygen and magnesium relative to iron that are larger in magnitude than those for thin-disk stars of the similar metallicity and are comparable to those for the oldest population in the Galaxy -- the halo. Similar behavior was found by Mashonkina and Gehren \cite{mash_eu} for the ratio of Eu to Ba. These results testify to a dominance of Type II supernovae in nucleosynthesis at the epoch of formation of the thick disk,
implying an old age for the thick disk, comparable to the age of the Galaxy, and therefore rapid formation of this structure. Other studies based on large stellar samples \cite{Bensby2005A&A...433..185B,2012A&A...545A..32A,Bensby2014} have confirmed differences in the chemical histories of the thin and thick disks of the Galaxy. As a rule, the halo is considered separately from the disks, 
and many studies have been dedicated to investigating the chemical enrichment of the halo based on observations of 
very metal-poor ($-4 \le$ [Fe/H] $\le -2$) stars; see, e.g., \cite{Cayrel2004,2009A&A...501..519B,Cohen2013,2013ApJ...762...26Y} and references therein).

There are very few studies in the literature in which elemental abundances are determined using common and accurate methods for representative samples, including stars of various populations in the Galaxy. One such study is that of Bergemann et al. \cite{2017ApJ...847...16B}, in which the [Mg/H] ratio was analyzed for stars of the halo, thick disk, 
and thin disk for metallicities $-2.5 \le$ [Fe/H] $\le -0.4$. The metallicity range $-1 \le$ [Fe/H] $\le 0$ was studied in \cite{Bensby2005A&A...433..185B,2012A&A...545A..32A,Bensby2014}. High-resolution spectra were obtained for $\sim 10^5$ stars in the Gaia-ESO Survey (see the description of this project in \cite{2013Msngr.154...47R} and the first results for $\alpha$-process elements in the thin and thick disks \cite{2014A&A...567A...5R}), and for $\sim 3 \cdot 10^5$ stars in the Apache Point Observatory Galactic Evolution Experiment (APOGEE \cite{2017AJ....154...94M}). In the vast majority of cases these stars had [Fe/H] $> -1$. The first published results for the Galactic Archaeology with HERMES (GALAH \cite{2018MNRAS.478.4513B}) project, which includes
primarily thin-disk and thick-disk stars, refer to metallicities in the range $-0.7 \le$ [Fe/H] $\le +0.5$ \cite{2019A&A...624A..19B}. A comparative analysis of the chemical compositions
of stars of the thick disk and halo with metallicities $-3.3 <$ [Fe/H] $< -0.5$ was carried out by Ishigaki et al. \cite{2012ApJ...753...64I}.

For chemical-evolution studies, it is important to have data on the chemical compositions of stars over a broad range of metallicity, and very important for these data to be uniform and accurate, since the evolutionary variations of various abundance ratios over the the Galaxy life do not exceed 0.4-0.5~dex.
It is desirable to detect not only variations, but also their time (Galactic epoch) dependence. In our
earlier studies, we selected two samples of stars, each of which were uniformly distributed in metallicity in 
the ranges $-2.6 \le$ [Fe/H] $ \le +0.3$ \cite{2015ApJ...808..148S,lick_paperII} and $-4 \le$ [Fe/H] $ \le -1.8$ \cite{dsph_parameters,2017A&A...608A..89M}, and used common methods to determine the atmospheric parameters and abundances of a large set of elements. This study was motivated by two circumstances.

\begin{itemize}
\item In \cite{2015ApJ...808..148S,lick_paperII}, the thick-disk population is represented by a small number of stars (7) in a narrow range of metallicity (from $-0.70$ to $-0.98$) and the star HD~94028 with [Fe/H] = $-1.47$. We have now increased the number of stars and expanded the metallicity range to
[Fe/H]  $\sim -0.15$.
\item In \cite{2015ApJ...808..148S} and \cite{dsph_parameters}, the surface gravities ($\lgg$) were determined based on ionization equilibrium of Fe~I/Fe~II and the spectral line formation without assuming LTE (a non-LTE, or NLTE, approach). These spectroscopic determinations require verification using the Gaia~DR2 parallaxes \cite{2018A&A...616A...1G}.
\end{itemize}

The aim of our current study is a comparative analysis of the abundances of the $\alpha$-process elements (Mg, Si, Ca, and Ti) relative to iron for stars of three populations: the thin disk, thick disk, and halo. We have verified the accuracy of the available data \cite{2015ApJ...808..148S,dsph_parameters} using the Gaia~DR2 parallaxes \cite{2018A&A...616A...1G}, and expanded the sample by adding 20 stars, predominantly of the thick disk. The atmospheric parameters and Mg,
Si, Ca, and Ti abundances of these stars have been determined using the same method as was applied for
the remaining sample.

The structure of the paper is as follows. The sample of stars, observational material, and atmospheric
parameters are presented in Section~\ref{Sect:stars}. Section~\ref{sect:abund} presents our determinations of the Mg, Si, Ca, and Ti abundances for the new sample stars. Section~\ref{sect:trends} 
presents our analysis of the data obtained and our conclusions are formulated in Section~\ref{sect:conclusions}.

\section{STELLAR SAMPLE, OBSERVATIONAL MATERIAL, ATMOSPHERIC PARAMETERS }
\label{Sect:stars}

\subsection{Stellar Sample, Identification of Galactic Population }

Our full sample contains 94 stars with metallicities $-4 \le$ [Fe/H] $ \le +0.3$. Of these, 51 stars are nearby ($d <$ 500~pc) dwarfs and subdwarfs from \cite{2015ApJ...808..148S,lick_paperII}, 23
are halo giants at distances up to 8~kpc from \cite{dsph_parameters,2017A&A...608A..89M}, and 20 have been added in our current study. These 20 stars were chosen using archival spectra of Fuhrmann \cite{Fuhrmann1998,Fuhrmann2004}. We gave preference to thick-disk stars with [Fe/H]  $> -0.7$. Note that all our studies have excluded variables, binaries with lines of both components in their spectra, and the carbon-enhanced stars. We imposed two additional requirements for halo giants. First, they must not have passed through a stage in which nucleosynthesis products are carried from the core to the atmosphere; i.e., their chemical compositions must reflect the chemical composition of the gas from which the star formed. Second, giants must have precise photometric magnitudes, so that the effective temperatures ($\Teff$) obtained using different color indices differ by no more than 100~K.

When determining membership of a star to a particular Galactic subsystem, the main criterion used was its kinematic characteristics. 
We calculated the vector Galactic velocities relative to the local
standard of rest ($U,V,W$) using a star's position, proper motion, parallax, and the zero velocity from
the Gaia~DR2 data \cite{2016A&A...595A...1G, 2018A&A...616A...1G}. Six stars (HD~22484, HD~30562, HD~49933, HD~84937, HD~106516, HD~114762)
are absent from Gaia~DR2, and we used data from the Hipparcos catalog for them \cite{2007A&A...474..653V}. 
Since the accuracy of the Gaia radial velocities is insufficient in some cases, we compared these values with those from the 
III/252 catalog \cite{2006AstL...32..759G}. If the accuracy of the III/252 data was higher and the difference between the two radial 
velocities exceeded 0.5~$\kms$, we adopted the III/252 values. The velocity of the Sun relative to the local standard of rest was taken to 
have coordinates (-10.2, +14.9, +7.8)~$\kms$ \cite{2005A&A...430..165F}. The probability of a star being a member of a specific Galactic subsystem
was calculated using formulas from \cite{2004A&A...418..551M}. We considered three subsystems: the thin disk, thick disk, and halo. Their fractional 
contributions to the total stellar population are 0.94, 0.0585, and 0.0015, their $V$ velocity shifts are $-15$, $-46$, and $-220$~$\kms$, 
and their
velocity dispersions are 35, 67, and 160~$\kms$ in $U$, 20, 38, and 90~$\kms$ in $V$, and 16, 35, and 90~$\kms$ in $W$. 
We used the Gaia~DR2 data to calculate the kinematic characteristics for both the 20 added stars (Table~\ref{tab:param}) and 
the sample of \cite{2015ApJ...808..148S}. Since these are nearby stars, the new velocities ($U,V,W$) are close 
to those calculated earlier using Hipparcos data \cite{2007A&A...474..653V}. The use of kinematic characteristics alone does 
not always enable an unambiguous determination of a star's membership in the thin or thick disk. 
We will discuss such stars in Section~\ref{sect:notes}.

\begin{table*}[htbp]
	\caption{Atmospheric parameters and $U$, $V$, $W$ components of the Galactic velocities of the studied stars.} 
	\label{tab:param} 
	\tabcolsep2.0mm
	\begin{center}
	\begin{tabular}{rccccrrrccc}
		\hline\hline
 &  &  &  &  &  &  &   & \multicolumn{3}{c}{probability (\%)} \\
\cline{9-11}
\multicolumn{1}{c}{HD} & $\Teff$, & $\lgg$ & [Fe/H] & $\xi_t$, & \multicolumn{1}{c}{$U$} & \multicolumn{1}{c}{$V$} & \multicolumn{1}{c}{$W$} & thin & thick & halo \\
\cline{6-8}
 & К &  &  & $\kms$ & \multicolumn{3}{c}{$\kms$} & \multicolumn{1}{c}{disk} & \multicolumn{1}{c}{disk} & \\
		\hline      
  3795 &  5475 & 3.85 &  -0.61 & 1.0 &  52.0 &  -90.4 &  40.7 & ~0 & 99 &  0  \\     
 10519 &  5740 & 4.01 &  -0.64 & 1.1 &  97.6 &  -76.1 &  34.9 & ~1 & 98 &  0  \\
 18757 &  5650 & 4.30 &  -0.34 & 1.0 &  71.8 &  -81.1 & -28.1 & ~5 & 94 &  0  \\
 32923 &  5710 & 4.03 &  -0.26 & 1.2 &  26.1 &  -23.6 &  28.4 & 97 & ~2 &  0 \\   
 40397 &  5550 & 4.39 &  -0.17 & 1.0 & 106.0 &  -92.2 & -37.1 & ~0 & 99 &  0 \\
 55575 &  5960 & 4.29 &  -0.37 & 1.2 &  80.0 &   -1.6 &  32.5 & 85 & 14 &  0 \\      
 64606 &  5280 & 4.63 &  -0.68 & 1.0 &  80.5 &  -64.7 &   3.7 & 48 & 51 &  0  \\  
 65583 &  5315 & 4.50 &  -0.68 & 0.8 &  13.3 &  -90.7 & -30.4 & ~4 & 95 &  0 \\ 
 68017 &  5615 & 4.41 &  -0.43 & 0.9 &  49.1 &  -59.7 & -41.1 & 28 & 71 &  0  \\
 69611 &  5940 & 4.17 &  -0.60 & 1.2 &  36.9 & -146.5 & -45.3 & ~0 & 91 &  8  \\
102158 &  5800 & 4.24 &  -0.47 & 1.1 & 113.0 & -116.6 &  12.3 & ~0 & 98 &  1  \\
112758 &  5260 & 4.54 &  -0.44 & 0.7 &  76.9 &  -35.9 &  18.0 & 88 & 11 &  0  \\
114762 &  5930 & 4.18 &  -0.71 & 1.2 &  78.6 &  -66.4 &  56.8 & ~0 & 99 &  0 \\   
132142 &  5100 & 4.47 &  -0.42 & 0.7 & 107.1 &  -54.7 &  19.0 & 25 & 74 &  0 \\
135204 &  5420 & 4.44 &  -0.13 & 0.9 &  85.3 &  -99.0 & -15.1 & ~0 & 99 &  0 \\                                                                                                        
144579 &  5250 & 4.49 &  -0.66 & 0.8 &  35.9 &  -58.4 & -18.4 & 83 & 16 &  0 \\
184499 &  5745 & 4.07 &  -0.54 & 1.2 &  64.8 & -161.2 &  58.4 & ~0 & 59 & 40 \\
201891 &  5900 & 4.29 &  -0.97 & 1.2 & -86.6 & -110.9 & -54.9 & ~0 & 97 &  2 \\ 
221830 &  5770 & 4.14 &  -0.41 & 1.2 &  68.1 & -113.6 &  63.0 & ~0 & 97 &  2 \\
222794 &  5600 & 3.90 &  -0.70 & 1.2 &  73.0 & -103.9 &  83.0 & ~0 & 95 &  4 \\
		\hline
	\end{tabular}
\end{center}
\end{table*}

\subsection{Observations}

A list of the 20 added stars is presented in Table~\ref{tab:param}. Spectra for these stars were obtained by Fuhrmann in 1995-2001 \cite{Fuhrmann1998,Fuhrmann2004} on the 2.2-m telescope of the Calar
Alto Observatory using the FOCES spectrograph. 
The spectral resolution for most of the stars was R $\simeq 60\,000$; for HD~184499, HD~201891, and HD~221830, R $\simeq$ 45\,000. 
In all cases, the signal-to-noise ratio (SNR) exceeded 100. Since the spectral range covered is 4500-6600\,\AA, we could not determine the oxygen abundance. The results of our earier studies were also based on high-resolution spectra: the 51 stars from \cite{2015ApJ...808..148S} were observed with R $\simeq 60\,000$ on the 3-m
Shane telescope (the Hamilton spectrograph) at the Lick Observatory (USA); for 11 stars from \cite{dsph_parameters} their spectra were taken from the CFHT/ESPaDOnS and VLT/UVES archives; and for the remaining stars we used the line equivalent widths from \cite{Cohen2013}.

\subsection{Atmospheric Parameters, Verification of Spectroscopic $\lgg$}

As in our earlier studies, we have made use here of photometric effective temperatures. Each of the
20 added stars has $\Teff$ values derived from the infrared flux method (IRFM)\cite{Casagrande2010,Casagrande2011}. We calculated $\lgg$ values using the Gaia~DR2 parallaxes\cite{2018A&A...616A...1G}. The stellar masses required for these calculations were derived by Fuhrmann \cite{Fuhrmann1998,Fuhrmann2004} based on evolutionary tracks, the $V$ magnitudes were taken
from the SIMBAD\footnote{http://simbad.u-strasbg.fr/simbad/} database, and the bolometric
corrections were taken from the tables of Alonso et al. \cite{Alonso1995}. The stellar masses are in the range (0.7-1) $M_\odot$, with uncertainties of 0.1~$M_\odot$. These mass uncertainties lead to uncertainties in  $\lgg$ of no more than 0.05~dex.

\begin{figure*}  
	\centering
\resizebox{170mm}{!}{\includegraphics{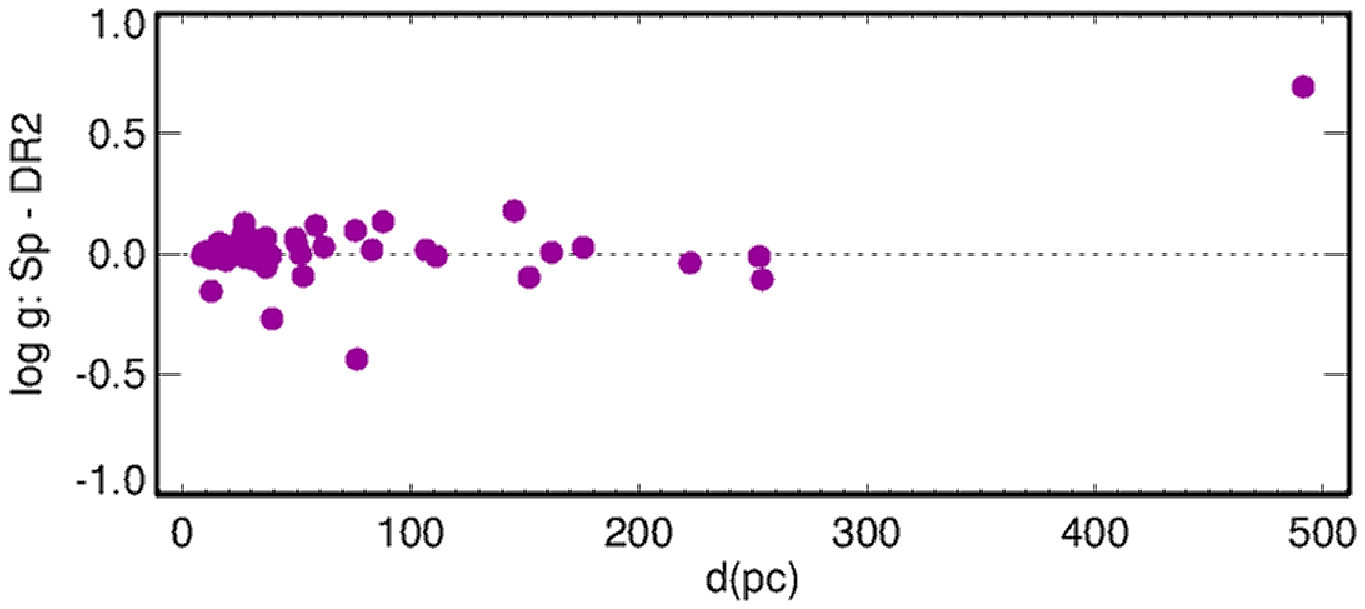}
	\includegraphics{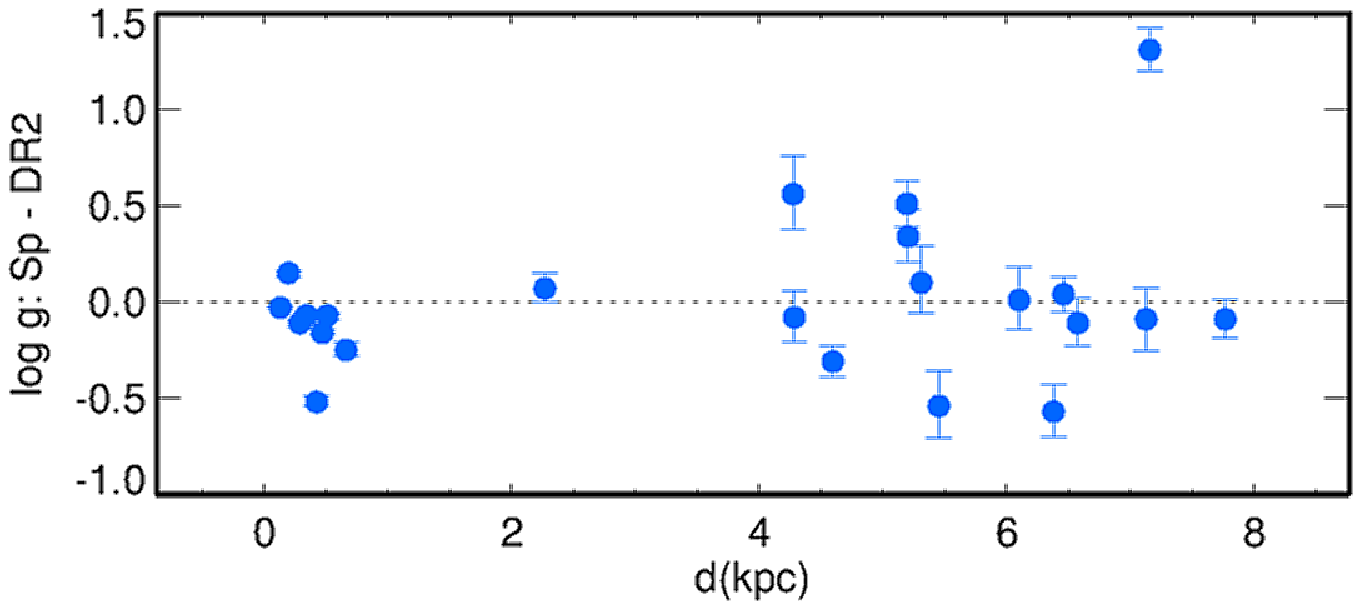}}
	\caption{Differences in the $\lgg$ values determined spectroscopically and using the Gaia~DR2 measurements as a function of the distances from \cite{2018AJ....156...58B}. The upper panel shows dwarfs and subgiants from \cite{2015ApJ...808..148S}, and the lower panel does halo giants from \cite{dsph_parameters}. The uncertainties in $\lgg$ correspond to the distance uncertainties presented in \cite{2018AJ....156...58B}. }
	\label{fig:dist}
\end{figure*}

Sitnova et al. \cite{2015ApJ...808..148S} and Mashonkina et al. \cite{dsph_parameters} determined spectroscopic gravities, $\lgg_{Sp}$, from the NLTE analysis of the Fe~I/Fe~II ionization equilibrium. Here, we verify these values by comparing them with the $\lgg_{DR2}$ values calculated using the Gaia DR2 parallaxes for nearby stars ($d <$ 500~pc) and the distances obtained in \cite{2018AJ....156...58B} based on Gaia DR2 measurements for more distant objects. The $V$ magnitudes were taken from SIMBAD and the bolometric corrections from the tables of Alonso et al. \cite{Alonso1995,Alonso1999}. The masses of the dwarfs and subgiants were determined by Sitnova et al. \cite{2015ApJ...808..148S} using evolutionary tracks, 
and the mass $M = 0.8 M_\odot$ was adopted for halo giants. We note that the masses of stars with [Fe/H] $\le -1$ have uncertainties no worse than 0.05~$M_\odot$, because it is known that these are old objects. Figure~\ref{fig:dist} shows that the spectroscopic $\lgg_{Sp}$ have a very good accuracy for both
dwarfs and giants. For the sample of dwarfs from \cite{2015ApJ...808..148S}, we obtained $\lgg_{Sp} - \lgg_{DR2} = 0.01\pm0.14$, with the omission of HD~138776 and BD~$-13 ^\circ3442$. Bringing into agreement the iron abundances from lines of two different ionization states with using $\lgg_{DR2}$ requires that $\Teff$ of HD~138776 be raised by roughly 300~K and $\Teff$ be decreased by a similar amount for BD~$-13 ^\circ3442$. For the latter star, our $\Teff$ value is based on the IR-flux method, which has yielded
similar values in various studies: $\Teff$ = 6364~K \cite{GH2009A&A...497..497G}, 6434~K \cite{Casagrande2010}, and 6442~K \cite{2005ApJ...626..446R}. There are no $\Teff$(IRFM) for HD~138776. Masana et al.  \cite{2006A&A...450..735M} give a photometric temperature of $\Teff$ = 5830$\pm$88~K, which is higher than our value by 180~K, but is not sufficiently high to provide Fe I/Fe II ionization equilibrium.

We obtained a good agreement between the spectroscopic and astrometric gravities for the giants, with the mean difference being $\lgg_{Sp} - \lgg_{DR2} = -0.05\pm0.13$ in cases when the uncertainty in $\lgg_{DR2}$ due to the distance uncertainty does not exceed 0.12~dex (10 stars). Among nearby ($\sim$500~pc) halo giants, a large difference between $\lgg_{Sp}$ and $\lgg_{DR2}$ was obtained for HD~8724. Bringing the
abundances determined from Fe~I and Fe~II lines into agreement requires raising $\Teff$ by roughly 300 K. 
We consider this to be unlikely, since independent determinations using the IR-flux method have yielded $\Teff$ = 4535~K \cite{Alonso1999}, 4540~K \cite{2005ApJ...626..446R}, and 4630~K \cite{GH2009A&A...497..497G}, which are all close to our value, $\Teff$ = 4560~K. The accuracies of the Gaia DR2 parallaxes worsen with distance, and it is likely that the distance uncertainties are higher than those given in \cite{2018AJ....156...58B}. At $d >$ 4~kpc, the observed scatter in $\lgg_{Sp} - \lgg_{DR2}$ cannot be due to uncertainties in the spectroscopy. The uncertainties in the spectroscopic method, for example, due to NLTE treatment, the use of classical homogeneous, plane-parallel model atmospheres, and the atomic data used for the iron lines, have a systematic, but not a random character.

We can draw the following conclusions from our comparison of the $\lgg_{Sp}$ and $\lgg_{DR2}$ values.

\begin{itemize}
\item For distant stars ($d >$ 4~kpc), the Gaia DR2 data do not provide the required accuracy in the
surface gravities. Spectroscopic methods remain irreplaceable for these objects.
\item For both dwarfs and giants of late spectral types, the NLTE analysis of the Fe~I/Fe~II
ionization equilibrium provides good accuracy in the $\lgg$ values. Therefore, we consider the
set of $\lgg$ values obtained for our full sample, partially from distances (20 nearby stars) and
partially spectroscopically (74 stars), to be uniform.
\item The Gaia DR2 data for the nearby ($d <$ 0.5~kpc) objects BD~$-13 ^\circ3442$, HD~8724, and possibly HD~138776 must be verified and updated.
\end{itemize}

We determined the iron abundances of the 20 stars using their Fe~II lines and assuming LTE, since the NLTE effects are small, when [Fe/H] $\ge -1$ \cite{mash_fe}. The atmospheric parameters derived are presented in Table~\ref{tab:param}. The microturbulence velocities $\xi_t$ were calculated using the empirical formula derived by Sitnova et al. \cite{2015ApJ...808..148S}.

\section{MAGNESIUM, SILICON, CALCIUM, AND TITANIUM ABUNDANCES}\label{sect:abund}

{\it Methods and programs.} 
We determined the Mg NLTE and LTE abundances for the 20 new stars using Mg~I lines, for Si using Si~I and Si~II lines, for Ca using Ca~I lines, and for Ti using Ti~II lines. In addition, we had to refine the Ca NLTE abundances for 51 stars from \cite{lick_paperII}. The reason this was necessary is that Zhao et al. \cite{lick_paperII} took into account Ca~I+H~I 
collisions in their NLTE computations in the theoretical approximation of Drawin \cite{Drawin1969} applying a scaling factor of $\kH$ = 0.1. 
The model atom for the Ca~I was later refined \cite{2017A&A...605A..53M} using the quantum-mechanical rate coefficients for Ca~I+H~I collisions 
from \cite{ca1_hydrogen}, and this model was then used in \cite{2017A&A...608A..89M} and in our current study when determining the Ca NLTE abundances. We used
the model atom for Mg~I from our earlier study \cite{mash_mg13}. A description and results of testing a new model atom for Si~I-II are being prepared for publication. The Ti abundance was determined using Ti~II lines under the LTE assumption, since NLTE effects are small for these lines in the studied range of stellar parameters \cite{2016AstL...42..734S}.

The system of the statistical equilibrium and radiative transfer equations in a specified model atmosphere was solved using the DETAIL program \cite{detail} with a modified opacity package, as described in \cite{mash_fe}.

\begin{table*}[htbp]
	\caption{List of lines, their atomic parameters, and the obtained solar abundances.} 
	\label{tab:lines} 
	\begin{center}
	\begin{tabular}{ccrccccccrccc}
		\hline\hline
  \ $\lambda$ & \Eexc & log$gf$ & log$C_6$ & \multicolumn{2}{c}{$\eps{}$} & \ \ & $\lambda$ & \Eexc & log$gf$ & log$C_6$ & \multicolumn{2}{c}{$\eps{}$} \\
\cline{5-6}
\cline{12-13}
 (\AA)     & (eV)    &         &          & {\scriptsize LTE} & {\scriptsize NLTE}  & &  (\AA) & (eV) & & & {\scriptsize LTE}  & {\scriptsize NLTE} \\                    
\hline
\multicolumn{6}{l}{~~Mg~I} & & \multicolumn{6}{l}{~~Ca~I } \\                                                                    
 4571.09 &  0.00 &  -5.62 & -31.74 &  7.67 &  7.70 & & 6166.44 &  2.51 &  -1.14 & -30.48 &  6.41 &  6.42 \\ 
 4702.99 &  4.34 &  -0.44 & -29.80 &  7.54 &  7.54 & & 6169.06 &  2.51 &  -0.80 & -30.48 &  6.43 &  6.41 \\
 4730.03 &  4.34 &  -2.35 & -29.89 &  7.80 &  7.81 & & 6169.56 &  2.53 &  -0.48 & -30.48 &  6.42 &  6.38 \\
 5528.41 &  4.34 &  -0.50 & -30.27 &  7.57 &  7.55 & & 6439.08 &  2.53 &   0.39 & -31.58 &  6.44 &  6.29 \\
 5711.07 &  4.34 &  -1.72 & -29.89 &  7.75 &  7.75 & & 6449.81 &  2.52 &  -0.50 & -31.45 &  6.48 & 6.26 \\ 
\multicolumn{6}{l}{~~Si~I}                         & & 6455.60 &  2.51 &  -1.34 & -31.45 &  6.38 &  6.31 \\                                                                  
 6155.13 &  5.62 &  -0.76 & -30.15 &  7.51 &  7.48 & & 6471.66 &  2.51 &  -0.69 & -31.58 &  6.46 &  6.31 \\
 6237.30 &  5.61 &  -0.98 & -30.15 &  7.43 &  7.40 & & 6493.78 &  2.52 &  -0.11 & -31.58 &  6.48 &  6.28 \\
\multicolumn{6}{l}{~~Si~II }                       & & 6499.65 &  2.51 &  -0.82 & -31.58 &  6.48 &  6.35 \\
 6347.11 &  8.12 &   0.17 & -31.55 &  7.71 &  7.62 & & \multicolumn{6}{l}{~~Ti~II}                      \\
 6371.37 &  8.12 &  -0.04 & -31.55 &  7.56 &  7.49 & & 5211.53 &  2.59 &  -1.41 & -31.82 &  4.95 &      \\
\multicolumn{6}{l}{~~Ca~I }                        & & 5336.79 &  1.58 &  -1.60 & -31.82 &  4.97 &      \\
 5512.98 &  2.93 &  -0.46 & -30.61 &  6.42 &  6.37 & & 5418.77 &  1.58 &  -2.13 & -31.82 &  5.00 &      \\
 5588.75 &  2.53 &   0.36 & -31.39 &  6.39 &  6.26 & & \multicolumn{6}{l}{~~Fe~II}                   \\ 
 5590.12 &  2.51 &  -0.57 & -31.39 &  6.36 &  6.34 & & 5425.26 &  3.20 &  -3.22 & -31.82 &  7.43 &      \\
 5857.45 &  2.93 &   0.24 & -30.61 &  6.44 &  6.35 & & 5991.38 &  3.15 &  -3.55 & -31.82 &  7.47 &      \\
 5867.57 &  2.93 &  -1.57 & -30.97 &  6.38 &  6.40 & & 6432.68 &  2.89 &  -3.57 & -31.82 &  7.44 &      \\
 6161.29 &  2.51 &  -1.27 & -30.48 &  6.37 &  6.39 & & 6456.38 &  3.90 &  -2.05 & -31.82 &  7.49 &      \\
 6162.17 &  1.90 &  -0.09 & -30.30 &  6.38 &  6.38 & &         &       &        & &       &      \\		
\hline
	\end{tabular}
\end{center}
\end{table*}

For both iron and $\alpha$-process elements, we used the same spectral lines (Table~\ref{tab:lines}) and atomic parameters as in our earlier study \cite{2015ApJ...808..148S}. We derived abundances using the synthetic-spectrum method; i.e., 
by fitting theoretical line profiles to the observations using the SynthV program \cite{2019ASPC....R} together with BinMag\footnote{http://www.astro.uu.se/\~{}oleg/binmag.html}. This software can compute theoretical NLTE spectra given so-called b factors (population
ratios obtained by solving the statistical-equilibrium equations and using the Boltzmann-Saha formulas),
which are computed in the DETAIL code for the levels in the model atom. The line lists used to compute
the synthetic spectra were taken from the VALD database of line parameters \cite{2015PhyS...90e4005R}.

{\it Model atmospheres} were obtained by interpolating in the MARCS\footnote{\tt http://marcs.astro.uu.se} model grid \cite{Gustafssonetal:2008} for specified $\Teff$/ $\lgg$/[Fe/H] values. We used the interpolation algorithm realized in the SME (Spectroscopy Made Easy) program \cite{2012ascl.soft02013V}.

{\it Analysis of solar lines.} 
As in \cite{lick_paperII}, we applied a differential approach to deriving abundances; i.e., for each individual line we substracted the solar abundance from the abundance obtained for a given star.
We determined the solar abundances for individual lines using spectra of the Sun as a star \cite{Atlas}. 
The MARCS model atmosphere with $\Teff$ = 5777~K and $\lgg$ = 4.44 was used, with a microturbulence velocity of $\xi_t$ = 0.9~$\kms$. The results are presented in Table~\ref{tab:lines}. We use a standard abundance scale in which $\eps{H}$ = 12.

\begin{table*}[htbp]
	\caption{NLTE abundances of Mg, Si, Ca, and Ti in the studied stars.} 
	\label{tab:abund} 
	\tabcolsep3.0mm
	\begin{center}
	\begin{tabular}{rcccccc}
		\hline\hline
\multicolumn{1}{c}{HD}  & [Fe/H] & [Mg/H] &[Si/H]$_{\rm I}$ & [Si/H]$_{\rm II}$ & [Ca/H] & [Ti/H] \\
		\hline      
   3795 &  -0.61 & -0.31(0.05) & -0.36(0.01) & -0.33(0.01) & -0.31(0.04) & -0.27(0.04) \\
  10519 &  -0.64 & -0.25(0.13) & -0.35(0.01) & -0.41(0.01) & -0.28(0.06) & -0.29(0.09) \\
  18757 &  -0.34 & -0.10(0.11) & -0.14(0.01) & -0.10(0.01) & -0.13(0.04) & -0.10(0.06) \\
  32923 &  -0.26 & -0.08(0.03) & -0.13(0.01) & -0.17(0.04) & -0.17(0.04) & -0.18(0.06) \\
  40397 &  -0.17 &  0.00(0.03) & -0.04(0.01) & -0.04(0.00) & -0.07(0.05) & -0.07(0.02) \\  
  55575 &  -0.37 & -0.24(0.07) & -0.26(0.01) & -0.31(0.03) & -0.28(0.05) & -0.22(0.05) \\
  64606 &  -0.68 & -0.59(0.12) & -0.53(0.04) & -0.41(0.13) & -0.59(0.08) & -0.52(0.04) \\
  65583 &  -0.68 & -0.41(0.07) & -0.44(0.04) & -0.35(0.02) & -0.39(0.04) & -0.46(0.01) \\
  68017 &  -0.43 & -0.15(0.03) & -0.24(0.03) & -0.18(0.02) & -0.23(0.03) & -0.18(0.04) \\
  69611 &  -0.60 & -0.20(0.11) & -0.32(0.04) & -0.41(0.01) & -0.28(0.05) & -0.24(0.07) \\
 102158 &  -0.47 & -0.15(0.06) & -0.20(0.04) & -0.23(0.08) & -0.15(0.12) & -0.15(0.03) \\
 112758 &  -0.44 & -0.15(0.07) & -0.28(0.03) & -0.26(0.03) & -0.13(0.06) & -0.23(0.02) \\
 114762 &  -0.71 & -0.48(0.09) & -0.42(0.02) & -0.47(0.06) & -0.53(0.06) & -0.44(0.03) \\
 132142 &  -0.42 & -0.13(0.05) & -0.21(0.01) & -0.10(0.04) & -0.15(0.04) & -0.20(0.05) \\
 135204 &  -0.13 &  0.08(0.02) & -0.02(0.01) &  0.01(0.01) &  0.01(0.04) &  0.00(0.03) \\
 144579 &  -0.66 & -0.39(0.09) & -0.43(0.03) & -0.30(0.03) & -0.42(0.05) & -0.41(0.02) \\
 184499 &  -0.54 & -0.25(0.07) & -0.25(0.02) & -0.24(0.08) & -0.32(0.08) & -0.23(0.04) \\
 201891 &  -0.97 & -0.78(0.11) & -0.78(0.01) & -0.75(0.04) & -0.79(0.16) & -0.74(0.03) \\
 221830 &  -0.41 & -0.14(0.07) & -0.18(0.04) & -0.21(0.02) & -0.20(0.06) & -0.10(0.07) \\
 222794 &  -0.70 & -0.41(0.07) & -0.42(0.04) & -0.37(0.03) & -0.44(0.05) & -0.37(0.07) \\
		\hline
\multicolumn{7}{l}{The rms uncertainties are given in parantheses.}		\\
\multicolumn{7}{l}{[Si/H]$_{\rm I}$ and [Si/H]$_{\rm II}$ are the abundances derived using the Si~I and Si~II lines, respectively.} \\
	\end{tabular}
\end{center}
\end{table*}

\begin{figure*}  
	\centering
\resizebox{170mm}{!}{\includegraphics{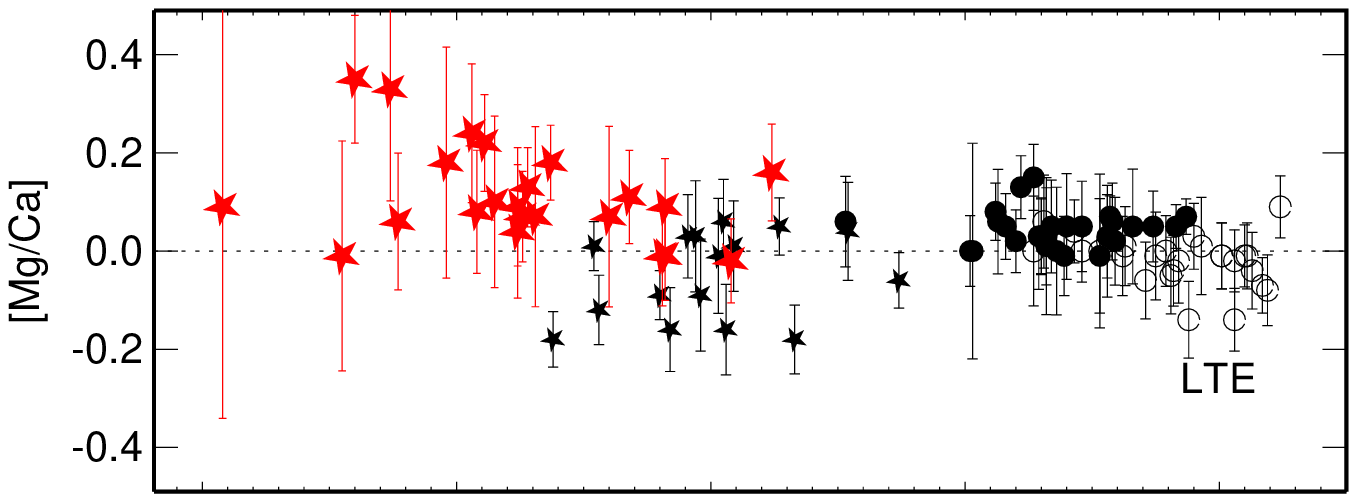}
	\includegraphics{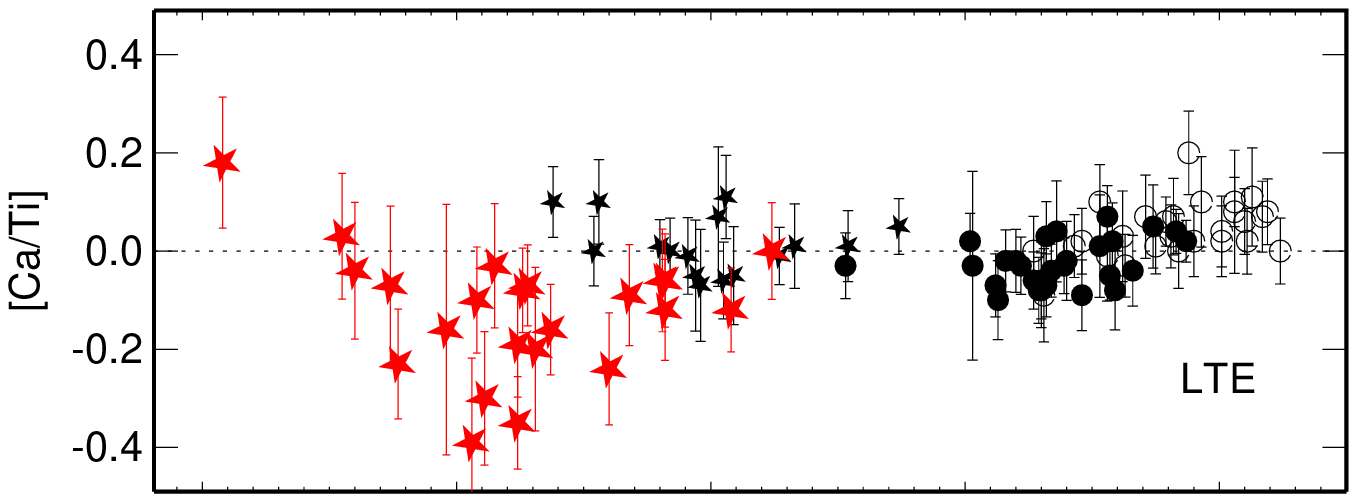}}

	\vspace{-7mm}	
\resizebox{170mm}{!}{\includegraphics{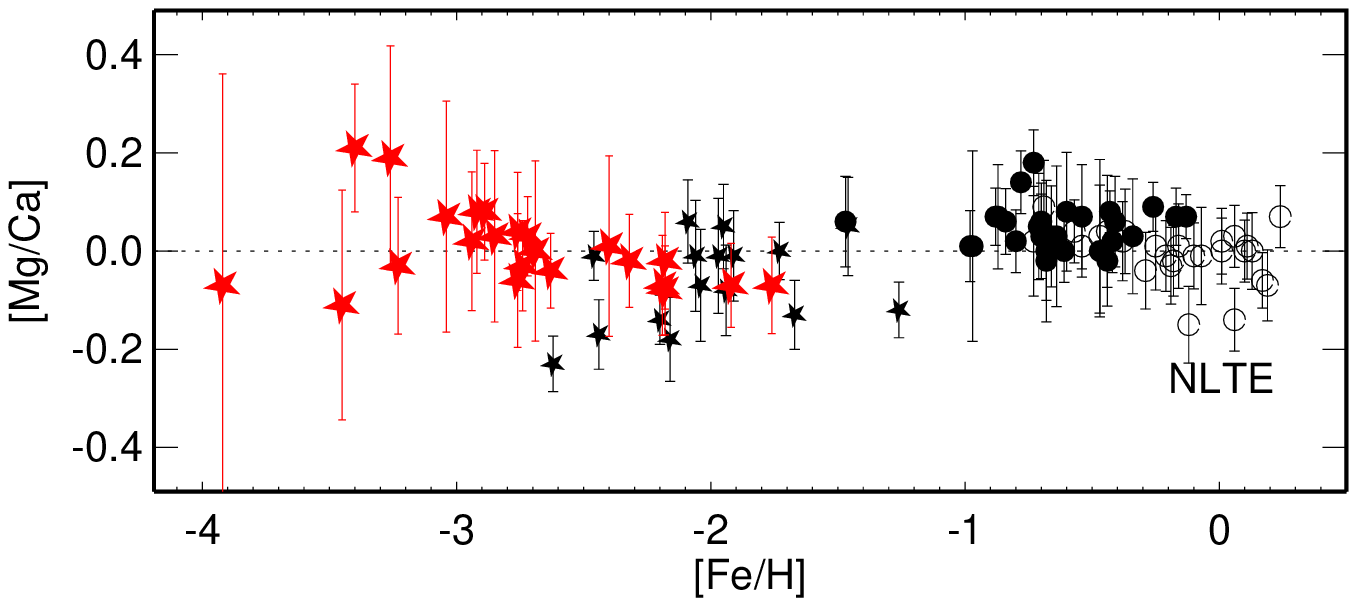}
	\includegraphics{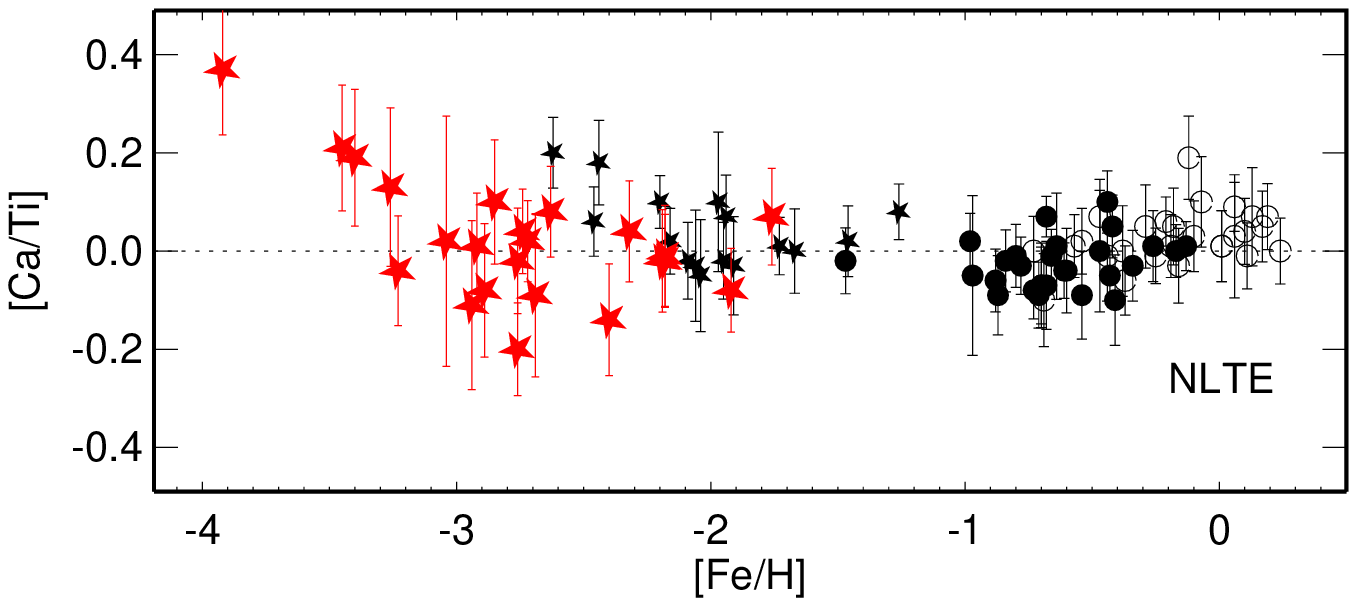}}
	\caption{The LTE (upper row) and NLTE (lower row) abundance ratios [Mg/Ca] (left) and [Ca/Ti] (right) for stars in the full sample. Thin-disk stars are marked by open circles, thick-disk stars by filled circles, halo dwarfs by small filled stars, and halo giants by large, red filled stars. }
	\label{fig:nlte}
\end{figure*}

Table~\ref{tab:abund} presents the mean differential abundances obtained in our NLTE computations. The accuracy of the mean abundance is given by the rms uncertainty$\sigma = \sqrt{\Sigma(\overline{x}-x_i)^2/(n-1)}$, where $n$ is the number of lines.

{\it Influence of NLTE effects on derived abundances.}
The difference between the NLTE and LTE abundances for individual lines is called the NLTE correction, $\Delta_{\rm NLTE} = \eps{\rm NLTE} - \eps{\rm LTE}$. The NLTE effects for Mg~I, Si~I, Si~II, and Ca~I for our new subsample of 20 stars depend only weakly on metallicity, in agreement with our earlier studies (see Fig.~2 in \cite{lick_paperII}). We can get an impression of the magnitude of $\Delta_{\rm NLTE}$ for various lines by analyzing the solar data (Table~\ref{tab:lines}). For Mg~I lines, $\Delta_{\rm NLTE}$s are small, from $-0.02$ to +0.03~dex. The Ca~I corrections have different signs and magnitudes for different Ca~I lines, ranging from $-0.22$ to +0.02 dex. We found negative $\Delta_{\rm NLTE}$ values for lines of both Si~I and Si~II, but with the corrections having larger magnitudes for
Si~II. Since the NLTE effects for each line are similar for the studied stars and for the Sun, the influence of these effects on the differential abundances [X/H] is low.

However, the situation is very different for a sample of stars with a wide range of metallicity. Figure~\ref{fig:nlte} shows [Mg/Ca] and [Ca/Ti] ratios obtained in our LTE and NLTE computations for the combined samples of Zhao et al. \cite{lick_paperII}, Mashonkina et al. \cite{2017A&A...608A..89M}, and
our current study. The difference between the LTE and NLTE values is small when [Fe/H] $ > -2$, but
this is not the case for the lower metallicity. In the LTE computations, [Mg/Ca] grows and [Ca/Ti]
falls with decreasing [Fe/H] for giants. This trend is removed for the NLTE values, and, on average, the
Mg/Ca and Ca/Ti ratios have their solar values over a broad interval of metallicity. In the case of Mg/Ca, this is true down to [Fe/H] $\simeq -4$, and for Ca/Ti down to [Fe/H] $\simeq -3.2$. It would be important to understand a reason for the growth of [Ca/Ti] when [Fe/H] $< -3.2$.

\section{[$\alpha$/Fe]--[Fe/H] DEPENDENCES }
\label{sect:trends}

\begin{figure*}  
	\centering
\resizebox{170mm}{!}{\includegraphics{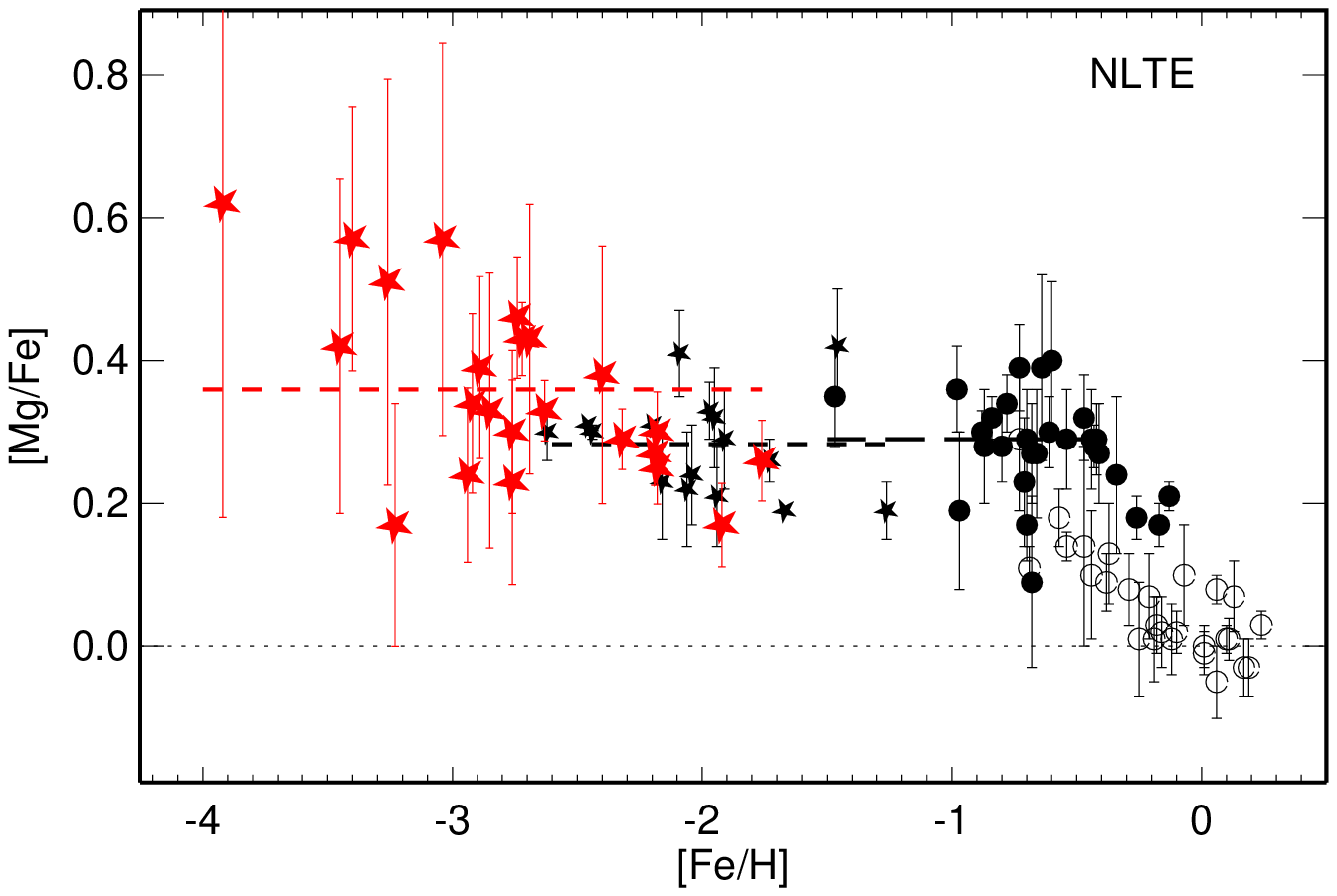}
	\includegraphics{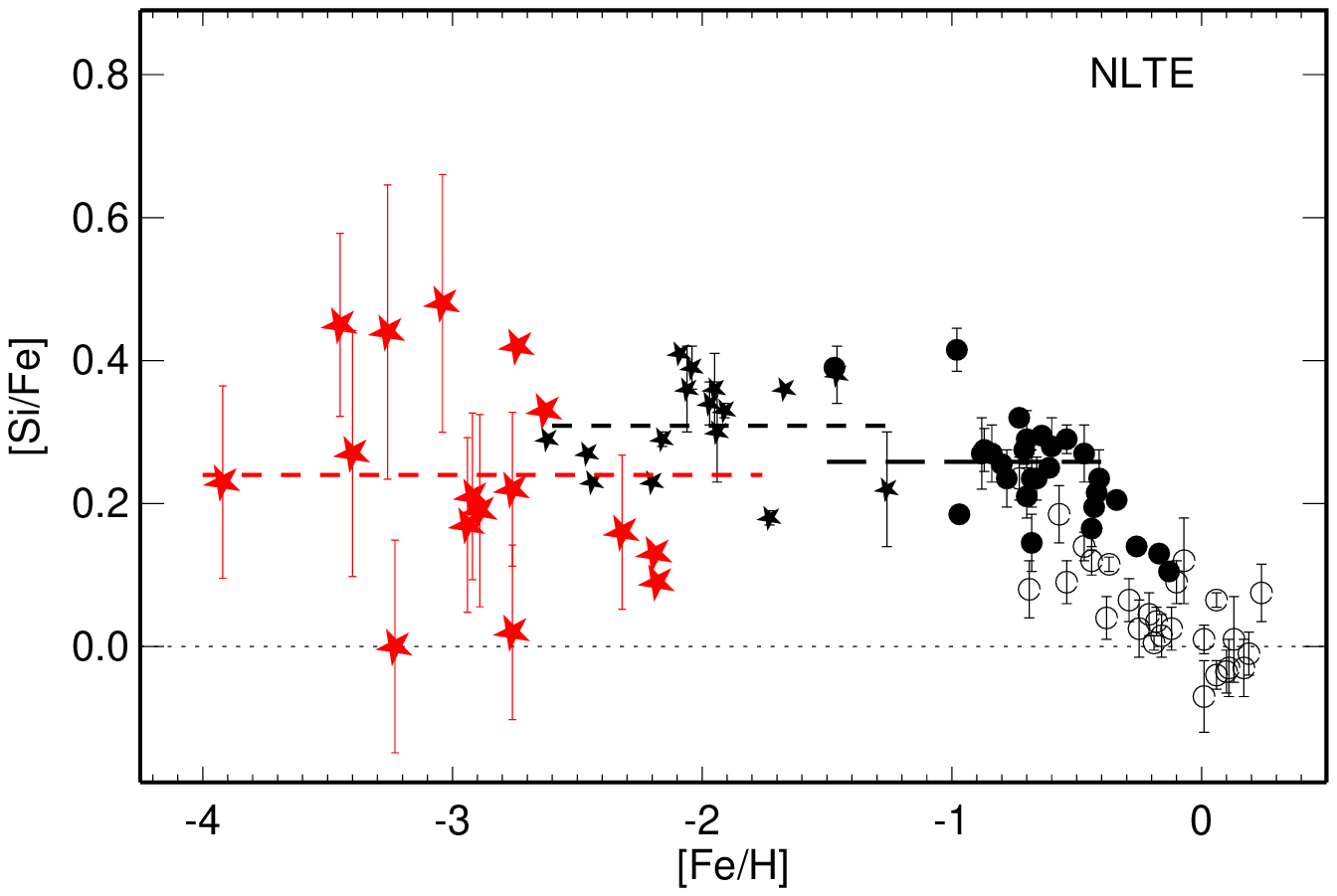}}
\resizebox{170mm}{!}{\includegraphics{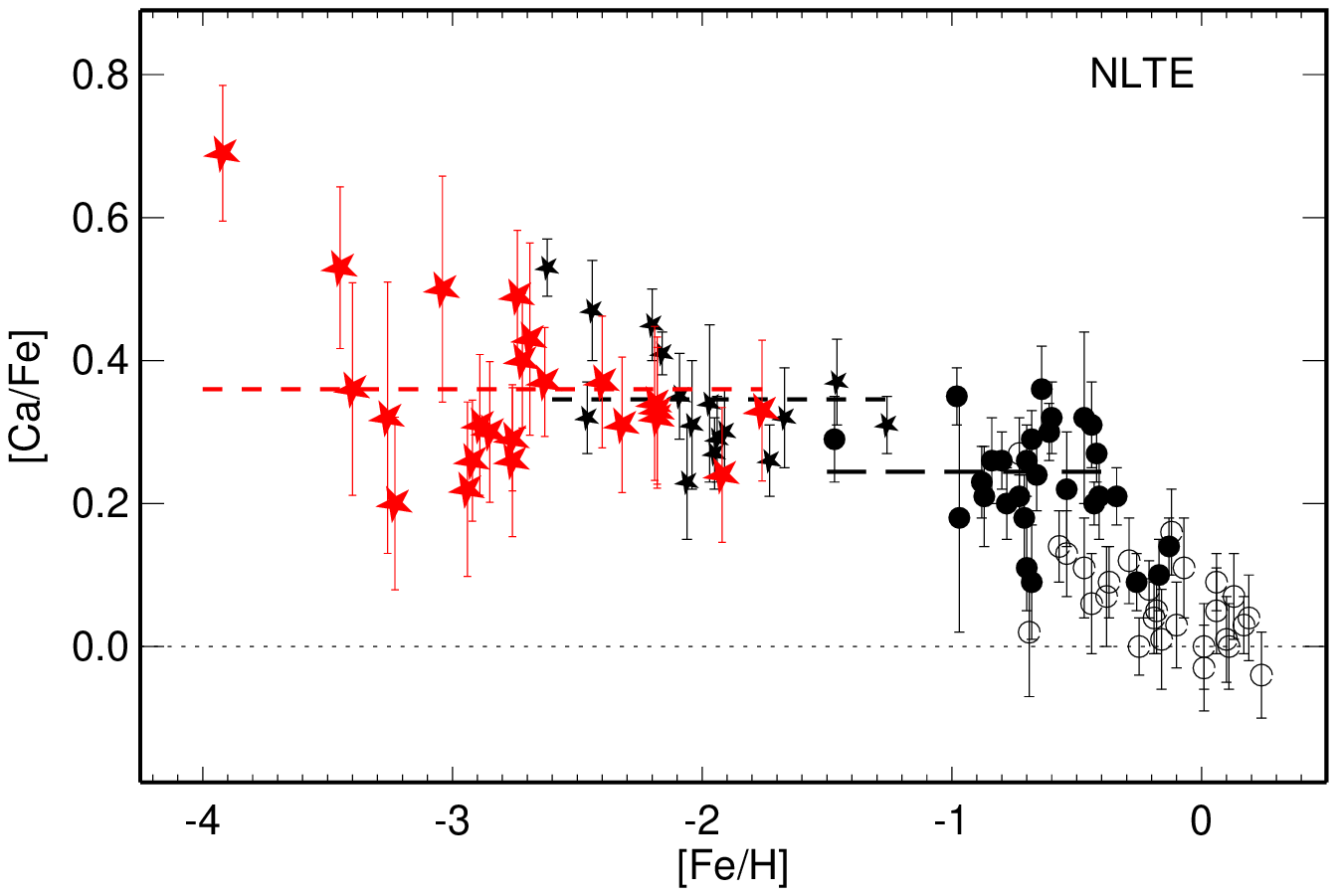}
	\includegraphics{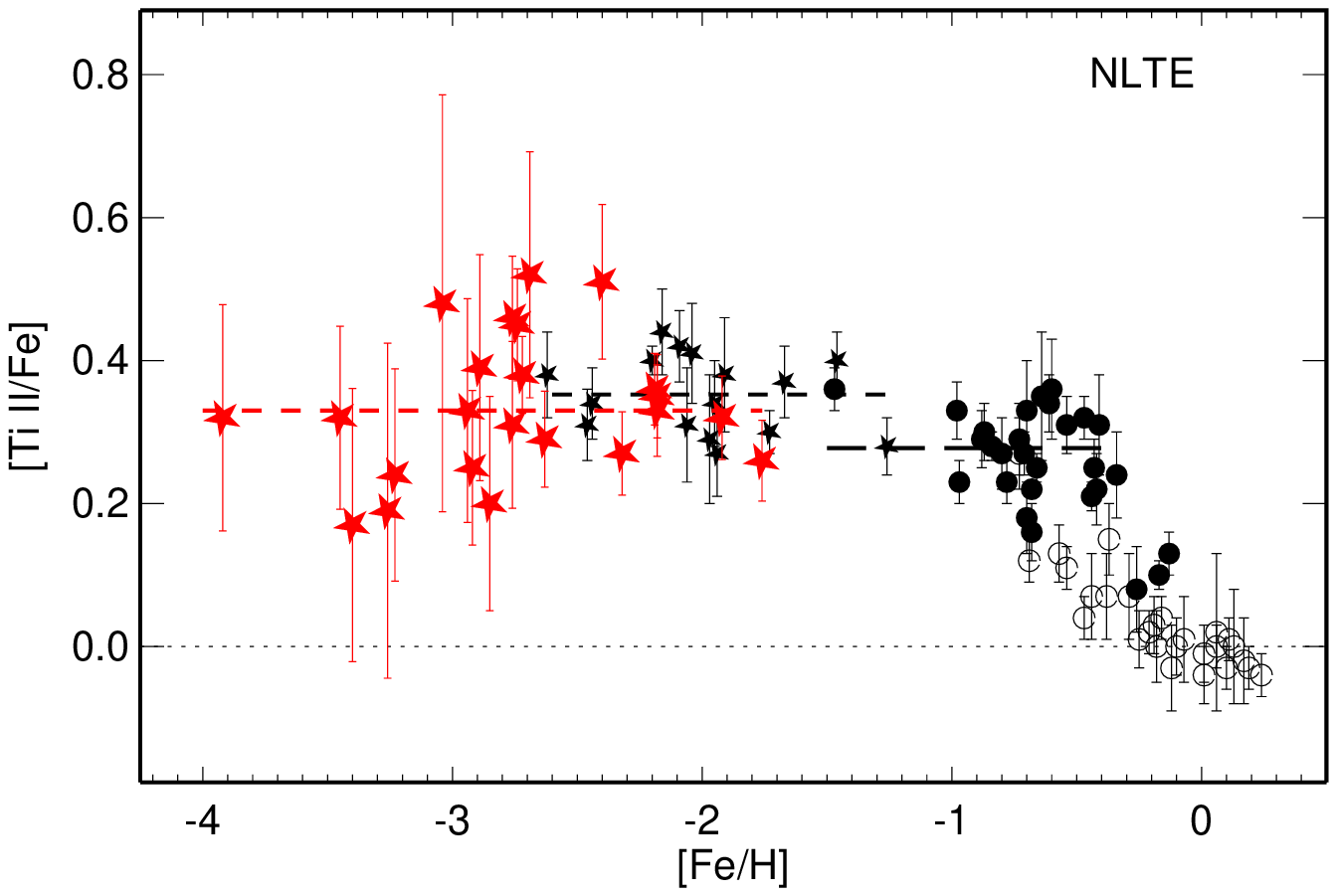}}
	\caption{The NLTE abundance ratios of $\alpha$-process elements to iron for stars of various Galactic populations. The symbols are the same as in Fig.~\ref{fig:nlte}. The dashed lines show the mean values for halo giants, halo dwarfs, and thick-disk stars with [Fe/H] $\le -0.4$.}
	\label{fig:alpha}
\end{figure*}

Figure~\ref{fig:alpha} presents ratios of the abundances of $\alpha$-process elements to iron for our full sample. The [Si/Fe] ratio was obtained by averaging the results for Si~I and Si~II lines. In addition to the abundances obtained in our current study, we also used the previously published results \cite{lick_paperII,2017A&A...608A..89M}. All the data were obtained using the same method without assuming LTE. The large scatter in [Si/Fe] for halo giants is due to the fact that only the Si~I 3905~\AA\ line could be used in this case, which is located in a spectral range with low SNR and strong blending, even when there is a strong deficit of iron.

\subsection{Thick and Thin Disk}\label{sect:notes}

Our sample includes 26 thin-disk stars with $-0.73 \le$ [Fe/H] $\le$ 0.24, and 28 thick-disk stars with
metallicities predominantly in the range $-0.98 \le$ [Fe/H] $\le -0.13$; one star, HD~94028 with [Fe/H] = $-1.47$, lies far from the others in the sample. The membership of stars to a particular population
was determined in \cite{2015ApJ...808..148S} based purely on its kinematics. Our experience leads us to be wary of this approach. We agree with the conclusions of Fuhrmann \cite{Fuhrmann1998,Fuhrmann2004},
Bensby et al. \cite{Bensby2014}, and Buder et al. \cite{2019A&A...624A..19B} that age is an important criterion for establishing membership of a star in the thin-disk or thick-disk population. Sitnova
et al. \cite{2015ApJ...808..148S} already discussed HD~59984, HD~105755, and HD~134169, which have low iron abundances but spatial velocities typical of the thin disk. These same authors present age estimates based on the evolutionary tracks of \cite{Yi2004}: $\sim 8$~billions years for the first two of these stars and $\sim 11$~billion years for HD~134169. Therefore, we have taken HD~59984 and HD~105755 to be associated with the thin disk and HD~134169 with the thick disk.

We estimated the ages of the 20 stars added to our sample based on the evolutionary tracks \cite{Yi2004}, and identified several objects that require comments.

\noindent
\underline{HD~40397 and HD~135204} have only small iron deficits, [Fe/H] = $-0.23$ and $-0.11$, 
respectively, but peculiar velocities typical of the thick disk, $V_{pec} = (U^2 + V^2 + W^2)^{1/2}$ = 145 and 132~$\kms$, respectively. 
Since both stars are far from leaving the main sequence ($\lgg$ = 4.39 and 4.44), their ages derived from the evolutionary tracks have large uncertainties. Fuhrmann \cite{Fuhrmann2004} notes them in his Fig.~34 as unclassified stars with intermediate chemistry and unknown age. Since our ages are estimated to be no less than nine billion years, we have taken them to be part of the thick-disk population.

\noindent
\underline{HD~112758 and HD~144579} belong to the thin disk with probabilities of 88 and 83~\%, respectively, but their ages are estimated to be 10 and 13.5 billion years. Therefore, we assigned them to the stellar population of the thick disk.

\noindent
\underline{HD~32923} has [Fe/H] = $-0.26$ and $V_{pec}$ = 45~$\kms$. Both of these values are typical for the thin disk, but its age is estimated to be 11 billion years. Fuhrmann \cite{Fuhrmann2004} assigned this star to a group of {\it transition stars} with intermediate chemistry and age. We have taken it to be a member of the thick disk population.

\noindent
\underline{HD~64606} can belong with similar probability to the thin disk and the thick disk population, according to adopted kinematic criteria (Table~\ref{tab:param}). This is a spectroscopic binary and seems to be a triple star, as discussed by Fuhrmann \cite{Fuhrmann2004}. Its parallax ($\pi_{\rm DR2} = 57.4339\pm0.4405$~mas) and proper motion were measured by Gaia with an accuracy better than 1~\%, therefore, it is unlikely that the uncertainty in identification of Galactic population is caused by the errors of astrometric measurements. The star position on the evolutionary tracks indicates a low mass of 
$\sim 0.7 M_\odot$, and the age can be between 5 and 10~Gyr. In contast to the other metal-poor stars, where $\alpha$-enhancements are similar for each of the four elements, Mg, Si, Ca, and Ti, HD~64606 reveals lower [Mg/Fe] = [Ca/Fe] = 0.09 compared with [Si/Fe] = 0.21. 
Based on large $V_{pec}$ = 103~$\kms$ and large eccentricity of the Galactic orbit of $e \simeq 0.75$, we assign HD~64606 to the thick disk population.

\begin{table*}[htbp]
	\caption{Mean NLTE abundance ratios of the $\alpha$-process elements to iron for different Galactic populations.} 
	\label{tab:pop} 
	\tabcolsep3.0mm
	\begin{center}
	\begin{tabular}{lcccccc}
		\hline\hline
  Population & [Fe/H] & N & [Mg/Fe] & [Si/Fe] & [Ca/Fe] & [Ti/Fe] \\
\hline
Halo giants & $-4.0$ to $-1.8$ & 23 & 0.36(0.13) &  & 0.36(0.11) & 0.33(0.10) \\
Halo dwarfs & $-2.6$ to $-1.2$ & 16 & 0.28(0.07) & 0.31(0.07) & 0.35(0.08) & 0.35(0.05) \\
Thick disk  & $-1.5$ to $-0.4$ & 24 & 0.29(0.07) & 0.26(0.06) & 0.24(0.07) & 0.28(0.06) \\
		\hline
	\end{tabular}
\end{center}
\end{table*}

We emphasize that we did not use any chemical properties of the stars when assigning them to particular Galactic populations. However, as in the pioneering study of Fuhrmann \cite{Fuhrmann1998} for [Mg/Fe]
and later studies for other $\alpha$-process elements (see, e.g., \cite{2012A&A...545A..32A,Bensby2014,2019A&A...624A..19B}), thick-disk and thin-disk stars display different behaviors of [$\alpha$/Fe] in the overlapping metallicity range (see Fig.~\ref{fig:alpha}), as is also true for
each of the elements individually: Mg, Si, Ca, and Ti. Namely, metal-poor thin-disk stars display
an excess of $\alpha$-process elements relative to iron, but [$\alpha$/Fe] falls off with growth in [Fe/H] starting from 0.29/0.24/0.19/0.27 for Mg/Si/Ca/Ti when [Fe/H] = $-0.73$ (HD~105755) to the solar values at the solar metallicity.

In the thick disk, the [$\alpha$/Fe] ratio displays different behaviors for [Fe/H] $\le -0.4$ and [Fe/H] $> -0.4$, although the number of stars is very low in the latter case (4). For stars with [Fe/H] $\le -0.4$, each of the four $\alpha$-process elements has a constant excess relative to iron at the level $\sim 0.3$ (see Table~\ref{tab:pop}), with only small star-to-star scatter. When [Fe/H] $> -0.4$, [$\alpha$/Fe] falls off with growing metallicity, testifying to the occurrence of Type~Ia supernovae at late stages of the formation of the thick disk ([Fe/H] $> -0.4$). This imposes constraints on the time scale of this epoch: its duration was longer than the time required for the occurrence of Type~Ia supernovae. 
Estimates of this latter time, as available in the literature, vary over very broad limits, from $\le$0.42~Gyr for prompt to $>$2.4~Gyr for delayed Type~Ia supernovae \cite{2012MNRAS.426.3282M}. Based on an analysis of [Eu/Ba] for thick-disk stars, Mashonkina et al. \cite{2003A&A...397..275M} showed that the thick disk formed between 1.1 and 1.6~Gyr after the beginning of the protogalactic collapse.

Our results for the thin disk agree with the earlier results not only qualitatively, but also quantitatively. However, our results for the thick disk differ from each of the earlier studies cited above. Fuhrmann \cite{Fuhrmann1998,Fuhrmann2004} did not have any thick-disk stars with [Fe/H] $> -0.4$, and obtained [Mg/Fe] $\simeq 0.4$ at the lower metallicities. Our results are in good ageement with those of Adibekyan et al. \cite{2012A&A...545A..32A} and Bensby et al. \cite{Bensby2014} for [Mg/Fe] and [Ti/Fe], but the [Si/Fe] and [Ca/Fe] ratios in both of those studies are $\sim 0.1$~dex lower than our values, and demonstrate a slope not only at [Fe/H] $> -0.4$, but also at lower [Fe/H]. We present a comparison with the results of \cite{2017ApJ...847...16B} in the following section.

\subsection{Thick Disk and Halo}

As has been shown in many earlier studies, stars of the halo and thick disk display an excess of $\alpha$-process elements relative to iron. Table~\ref{tab:pop} presents the mean abundance ratios [Mg/Fe], [Si/Fe], [Ca/Fe], and [Ti/Fe] for three subsamples of stars from our study: halo giants, halo dwarfs (and subgiants), and thick-disk stars with [Fe/H] $\le -0.4$. The ratio [Si/Fe] for the halo giants is an exception; for these stars, we did not calculate the mean due to the large scatter between individual stars (see Fig.~\ref{fig:alpha}). Leaving aside [Si/Fe], Fig.~\ref{fig:alpha} shows that the scatter
in [$\alpha$/Fe] about the mean values is nearly a factor of two greater at metallicities [Fe/H] $\le -2.6$, than for stars with higher [Fe/H] values. Of course, these are more distant objects with more uncertain interstellar absorptions and lower quality spectra, leading to higher uncertainties in $\Teff$, $\lgg$, and elemental abundances. However, we also cannot rule out chemical inhomogeneity of the gas at the epoch of formation of stars with [Fe/H] $< -2.6$. For example, the scatter decreases appreciably in this range if we consider ratios between $\alpha$-process elements (Fig.~\ref{fig:nlte}),
and not [$\alpha$/Fe].

We have obtained the following new results.
 
\begin{itemize}
\item The halo stars have similar [$\alpha$/Fe] values, independent of whether they are located in the
solar neighborhood (dwarfs) or at larger distances (most of the giants, see Fig.~\ref{fig:dist}). This testifies to a universal character of the evolution of the abundances of $\alpha$-process elements
in different regions and volumes of the Galaxy.
\item The halo stars display excesses relative to iron that are the same for all four $\alpha$-process
elements -- Mg, Si, Ca, and Ti -- at the level $\sim 0.3$. This places important observational
constraints on nucleosynthesis models.
\item Thick-disk stars with [Fe/H] $\le -0.4$ display the same excess of $\alpha$-process elements as halo stars. We note the lower ratio [Ca/Fe] = 0.24 for thick-disk stars compared to [Ca/Fe] = 0.35
for halo stars, but this difference lies within 1.5$\sigma$.
\end{itemize}

Let us compare our results with data from the literature. The study of Bergemann et al. \cite{2017ApJ...847...16B} is the only previous analysis for which NLTE abundances were determined, but this was done only for [Mg/Fe]. For their sample of thick-disk stars with $-2.1 <$ [Fe/H] $< -0.6$, they obtained the constant value [Mg/Fe] $\sim 0.3$, as we have in our current study, in spite of their use of $\ensuremath{\langle\mathrm{3D}\rangle}$ model atmospheres, which were made from 3D models by averaging over a computational box and over time. For halo stars with $-2.5 <$ [Fe/H] $< -2$, both 1D-NLTE and $\ensuremath{\langle\mathrm{3D}\rangle}$-NLTE approaches yielded the same relative abundances, [Mg/Fe] $\sim 0.3$ \cite{2017ApJ...847...16B}, again in agreement with our NLTE values. It is not clear why the  1D-NLTE and $\ensuremath{\langle\mathrm{3D}\rangle}$-NLTE results differ so strongly for halo stars with [Fe/H] $> -2$ (Fig.~5 in \cite{2017ApJ...847...16B}).

The sample of thick-disk stars of Ishigaki et al. \cite{2012ApJ...753...64I} includes 11 stars, most of which (8) have metallicities $-1.5 <$ [Fe/H] $< -0.6$, overlapping with our sample. All their abundances were obtained assuming LTE. The mean values presented in Table~8 of \cite{2012ApJ...753...64I} ([Mg/Fe]/[Si/Fe]/[Ca/Fe]/[Ti/Fe] = 0.32/0.32/0.23/0.37, thick disk, [Fe/H] $> -1.5$) are in good agreement with our results. Ishigaki et al. \cite{2012ApJ...753...64I} present separate abundances derived from Ti~I and Ti~II lines. As expected, the Ti~I abundances are lower than the Ti~II abundances, by 0.18~dex. We
used the Ti~II abundances for our comparison, since NLTE effects are negligible for these lines in this
metallicity range \cite{2016AstL...42..734S}. The agreement of our NLTE results with the LTE results of \cite{2012ApJ...753...64I} for [Mg/Fe], [Si/Fe], and [Ca/Fe] is not surprising, since the NLTE effects for Mg~I, Si~I, and Ca~I lines depend only weakly on metallicity when [Fe/H] $> -1.5$, and the differential approach minimizes deviations from LTE in the [X/Fe] values.

Ishigaki et al. \cite{2012ApJ...753...64I} note that they obtained lower [$\alpha$/Fe] values for halo stars than those for the thick disk, on average, and found a large scatter between individual stars. We have carefully analyzed Table~8 of \cite{2012ApJ...753...64I}. Indeed, in the entire metallicity range $-3.2 <$ [Fe/H] $< -0.6$, the rms uncertainties are 0.11 to 0.14~dex for various ratios, a factor of 1.5-2 higher than we obtained for halo stars with [Fe/H] $> -2.6$. For halo stars with [Fe/H] $> -1.5$, they indeed found lower [Mg/Fe] and [Si/Fe] values than those for the thick disk, by 0.17 and 0.12~dex, respectively, however, their [Ca/Fe] and [Ti/Fe] values agree within the uncertainties. Since our sample does not contain halo stars with [Fe/H] $> -1.5$, we cannot either confirm or refute their results. For halo stars with lower iron abundances, Ishigaki et al. \cite{2012ApJ...753...64I} found 
[Mg/Fe]/[Si/Fe]/[Ca/Fe]/[Ti/Fe] = 0.24/0.30/0.28/0.36, in agreement with the corresponding ratios for
the thick disk, and also consistent with our results within the uncertainties.

\begin{figure*}  
	\centering
\resizebox{170mm}{!}{\includegraphics{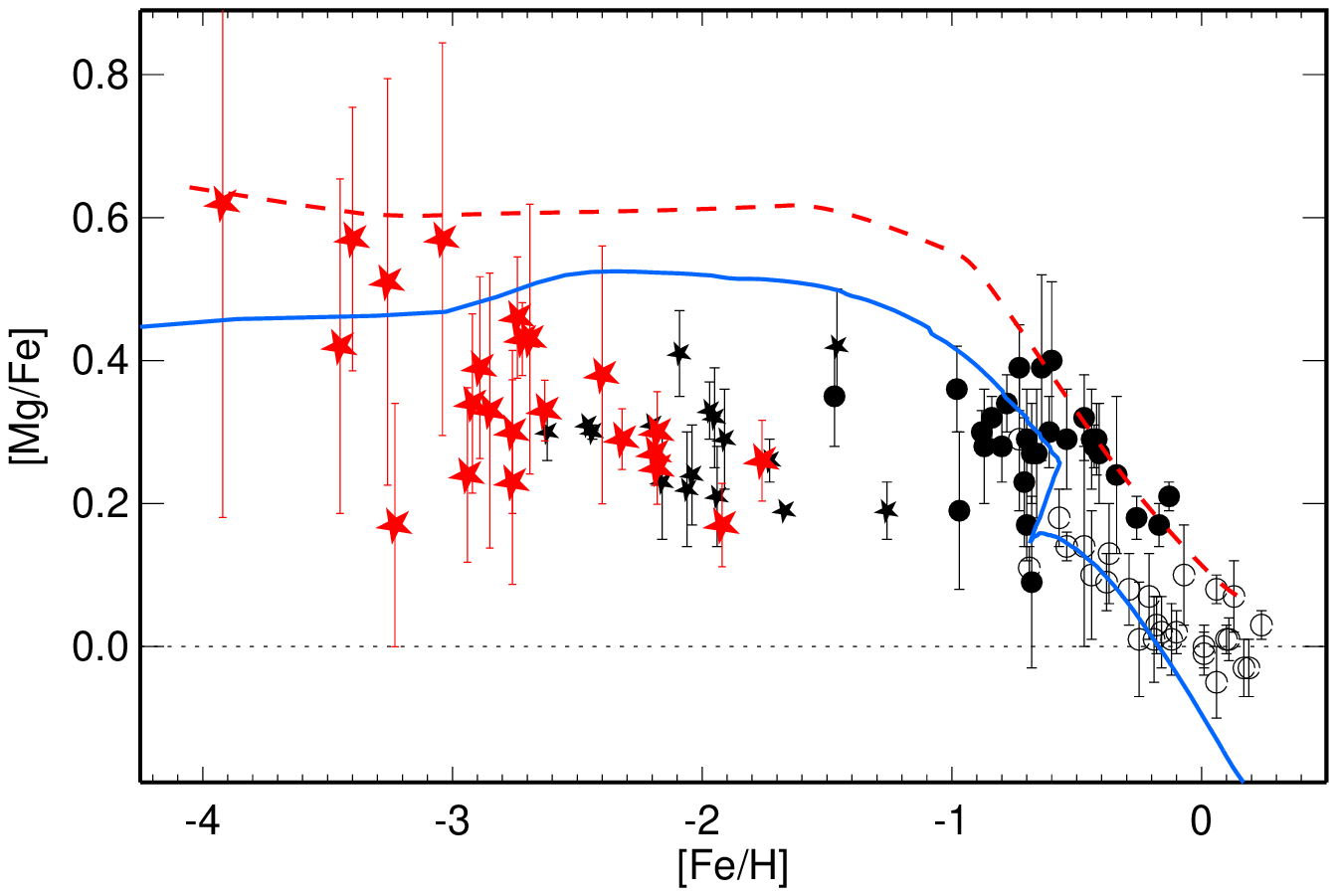}
	\includegraphics{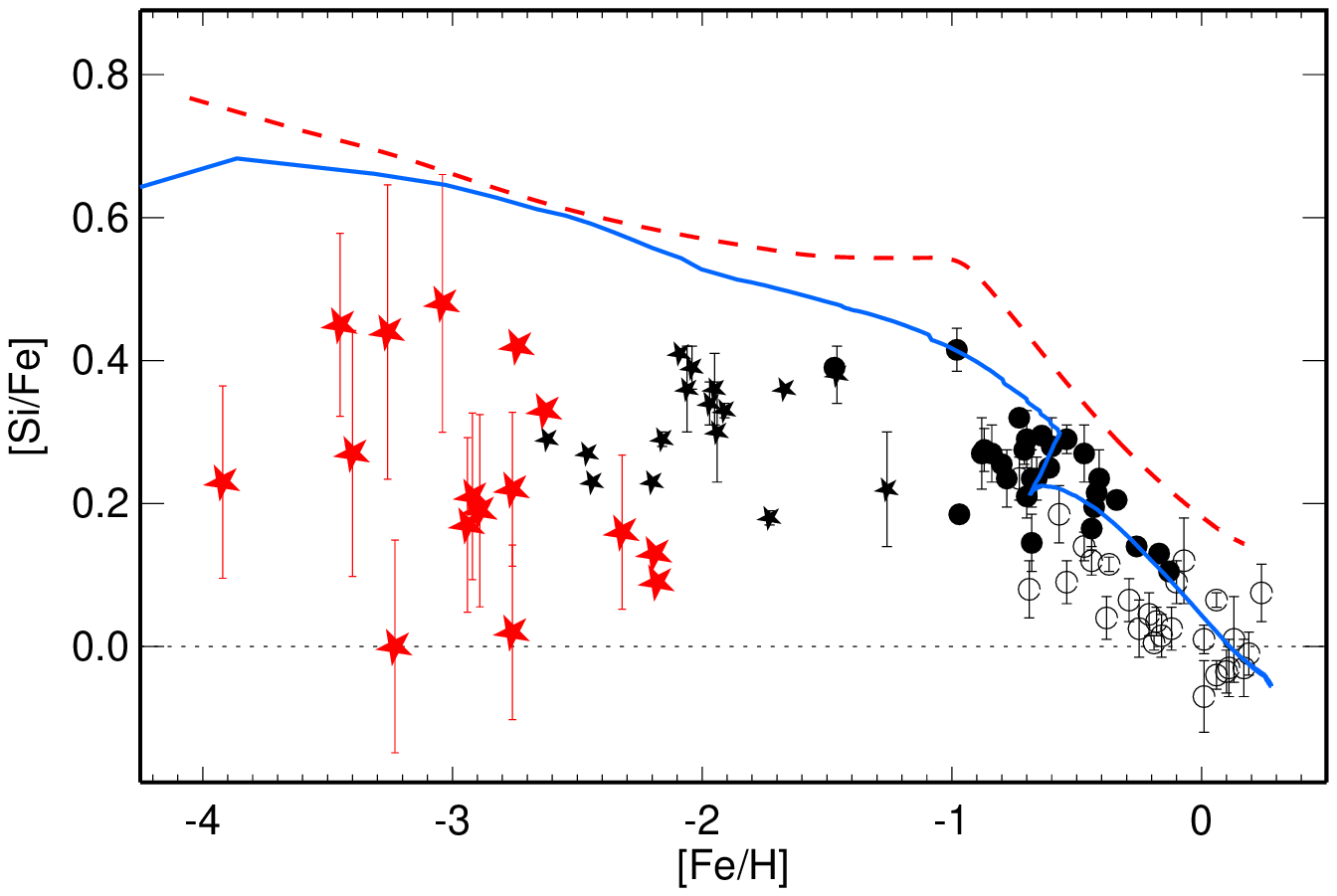}}
\resizebox{170mm}{!}{\includegraphics{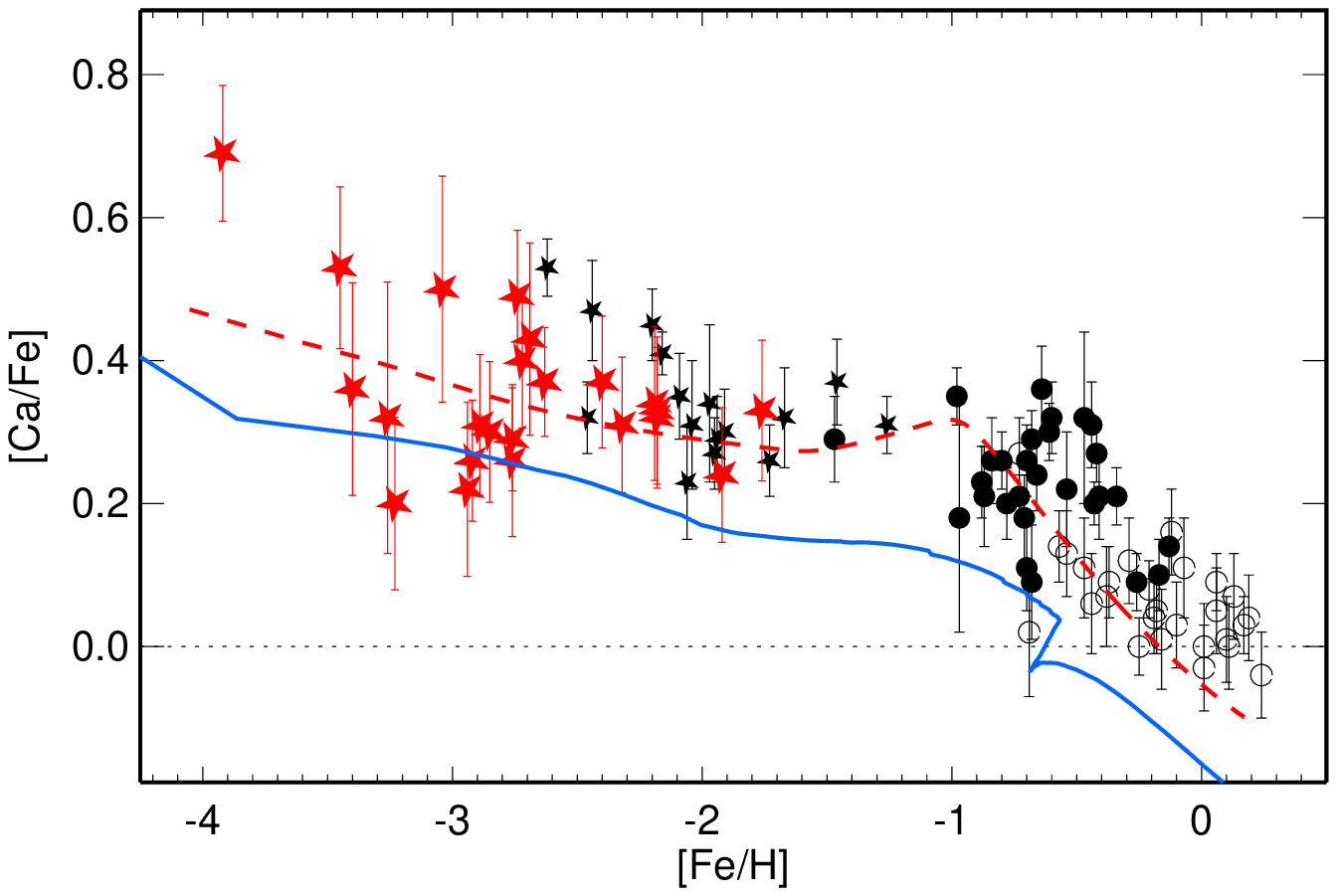}
	\includegraphics{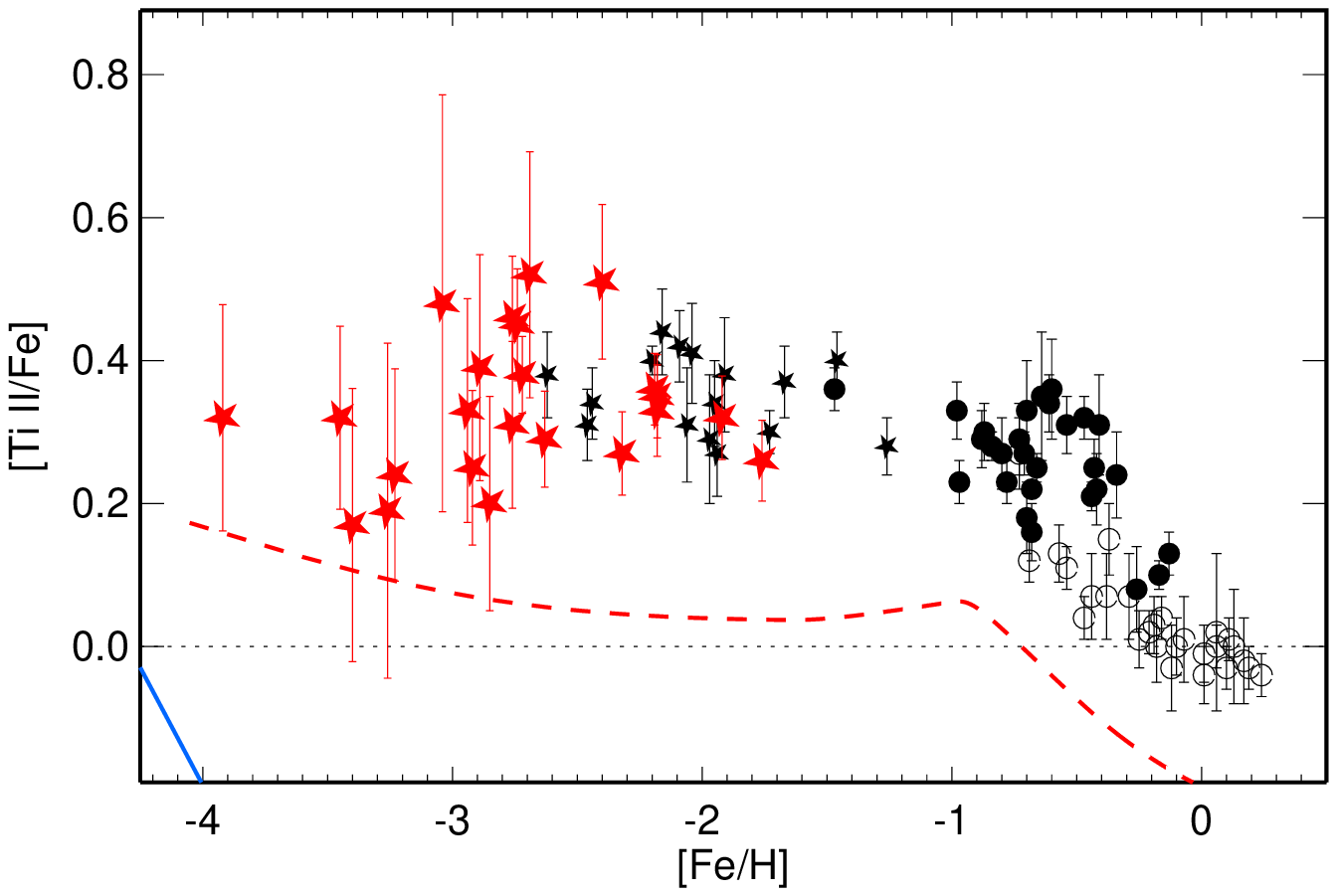}}
	\caption{Observed [$\alpha$/Fe] ratios compared to the Galactic chemical evolution models K15 (\cite{2016ApJ...817...53S}, dashed curve) and R10 (\cite{2010A&A...522A..32R}, solid curve). Symbols are the same as in Fig.~\ref{fig:nlte}.}
	\label{fig:gce}
\end{figure*}

\subsection{Comparison with Models for the Chemical Evolution of the Galaxy}

The uniformity and accuracy of the abundance ratios we have obtained over a large interval of metallicity
for a representative sample of 94 stars suggests that these results may be useful for studies of the
chemical evolution of Galactic populations. For a comparison of the observations and theory, we selected
two models for the chemical evolution of the Galaxy that are most often applied in the literature.
These are by Romano et al. \cite{2010A&A...522A..32R} (hereafter, R10) and by C. Kobayashi (hereafter, K15), as presented in \cite{2016ApJ...817...53S}. 
 These models assume different scenarios for the formation of the Galaxy, but include common sources of chemical enrichment of the interstellar
medium -- hypernovae, Type~II supernovae, Type~Ia supernovae, and the asymptotic giant branch stars, 
and for most elements use the same sources for the nucleosynthesis yields.
 
 As is shown in Fig.~\ref{fig:gce}, the R10 and K15 models correctly predict the [Mg/Fe], [Si/Fe], and [Ca/Fe] variations with [Fe/H]; i.e., 
enhancements of Mg, Si, and Ca for [Fe/H] $\le -1$ and the fall in [$\alpha$/Fe] with growth in [Fe/H] at higher metallicity.
This behavior indicates that Type~II supernovae dominated in nucleosynthesis until the iron abundance in
the Galaxy reached values [Fe/H] $\sim -1$, after which the occurrence of Type~Ia supernovae led to an increase
in the rate of production of iron and a drop in [$\alpha$/Fe]. However, neither model reproduces the
observational data quantitatively. The K15 model for [Ca/Fe] is closest to the observations, but that
model cannot fit the trends in the abundance ratios for the thick-disk and thin-disk stars, since it assumes
a one-zone model for the Galaxy. The R10 model allows for different chemical histories of the thick and
thin disks, but underestimates [Ca/Fe] over the entire range of metallicity. Both models predict very high Si enhancements
 for [Fe/H] $< -1$, but, as is discussed in \cite{lick_paperII},
the discrepancy with observations can be decreased by varying the masses of the pre-supernovae and the
energy of the Type~II supernova outbursts.
 
The most serious problems posed for the theory are the observations of [Mg/Fe] for [Fe/H] $< -1$ and of [Ti/Fe] for any [Fe/H]. The theory predicts that
Mg and O are synthesized in the same regions in the pre-supernova stage, so that they should have the
same nucleosynthesis yields (relative to the solar values); i.e., we would expect [Mg/Fe] = [O/Fe]
for metallicities [Fe/H] $< -1$. However, this is not confirmed by the observations: the [O/Fe] ratio is
much higher than [Mg/Fe] for halo stars, by 0.3~dex (see, e.g., \cite{2018AstL...44..411S}). As concerns Ti, in the theory, it is
synthesized as an iron-group element, and the model therefore does not predict an excess relative to iron;
however, Ti is manifest observationally in the same way as the $\alpha$-process elements.

\section{CONCLUSIONS}\label{sect:conclusions}

We have extended to 94 stars our sample of stars with atmospheric parameters and abundances of the $\alpha$-process elements Mg, Si, Ca, and Ti determined
using common and accurate methods. We draw the following methodical conclusions.

\begin{enumerate} 
\item Our comparison of $\lgg_{DR2}$, based on Gaia DR2 measurements, and $\lgg_{Sp}$, determined
spectroscopically from the Fe I/Fe II ionization eqilibrium method, has shown that the spectroscopic method
yields reliable results when the assumption of LTE is rejected.
\item For distant stars ($d >$ 4~kpc), the Gaia DR2 measurements do not provide the accuracy required
for reliable determination of their surface gravitaties.
\item We have confirmed the conclusion of earlier studies that the use of kinematic criteria alone does
not enable unambiguous determination of the membership of stars in specific Galactic populations. Stellar
ages are required in this case, if they can be estimated.
\end{enumerate} 

Our stellar sample covers metallicities $-4 <$ [Fe/H] $< 0.3$ and includes representative samples of thin-disk,
thick-disk, and halo stars. The uniformity and high accuracy of the data make them valuable for studies
of the chemical evolution of the Galaxy. We have confirmed the conclusions of earlier studies concerning
enhancements of Mg, Si, Ca, and Ti relative to Fe for halo and thick-disk stars, and the larger enhancements of these elements for thick-disk stars
compared to thin disk stars of close metallicity. We have also obtained the following new conclusions.

\begin{enumerate} 
\item In the thick disk, [Mg/Fe], [Si/Fe], [Ca/Fe], and [Ti/Fe] all remain roughly constant at $\sim 0.3$
 when [Fe/H] $\le -0.4$, and fall off at higher metallicities, indicating the onset of the
production of iron in Type~Ia supernovae.
\item Halo stars have the same [$\alpha$/Fe] ratios independent of whether they are located in the solar
neighborhood or at larger distances, up to $d \sim$ 8~kpc, testifying to a universal evolution of the abundance of $\alpha$-process
elements in various regions and volumes of the Galaxy.
\item For [Fe/H] $\le -2.6$, the scatter in [$\alpha$/Fe] increases, while the scatter of the abundance ratios between 
 different $\alpha$-process elements is small, possibly indicating incomplete mixing of nucleosynthesis
products at the epoch of formation of these stars.
\item For halo stars, the enhancements of each of the four $\alpha$-process elements Mg, Si, Ca, and Ti relative to iron 
are, on average, the same, at the level $\sim 0.3$~dex.
\item Thick-disk stars with [Fe/H] $\le -0.4$ have the same enhancements of $\alpha$-process elements as halo
stars.
\end{enumerate} 

We hope that these observational data will be used to refine models for nucleosynthesis and the chemical
evolution of the Galaxy.

{\it Acknowledgements.} We thank Klaus Fuhrmann for kindly presenting us with spectra obtained with the FOCES spectrograph on the 2.2~m telescope of
the Calar Alto Observatory. This study has made use of the ADS\footnote{http://adsabs.harvard.edu/abstract\_service.html}, 
SIMBAD, MARCS, and VALD databases.

\end{document}